\documentclass[twocolumn,superscriptaddress, longbibliography,
amsmath,amssymb,
]{revtex4-2}

\usepackage{graphicx}
\usepackage{dcolumn}
\usepackage{bm}
\usepackage{xcolor}

\DeclareMathOperator{\Rey}{\mathit{Re}}
\DeclareMathOperator{\Rm}{\mathit{Rm}}
\DeclareMathOperator{\Pm}{\mathit{Pm}}
\DeclareMathOperator{\Po}{\mathit{Po}}

\begin{document}
	
	\preprint{APS/123-QED}
	
	\title{Magnetic Dynamo Driven by Inertial Waves }
	
	\author{Ashish Mishra}
	\email{a.mishra@hzdr.de}
	\affiliation{Helmholtz-Zentrum Dresden-Rossendorf, Bautzner Landstr. 400, D-01328 Dresden, Germany}
	\author{George Mamatsashvili}%
	\affiliation{Helmholtz-Zentrum Dresden-Rossendorf, Bautzner Landstr. 400, D-01328 Dresden, Germany}
	\affiliation{Abastumani Astrophysical Observatory, Abastumani 0301, Georgia}
	\author{Michael Le Bars}
	\affiliation{Aix Marseille Univ, CNRS, Centrale Marseille, IRPHE UMR 7342 - Marseille, France}
	\author{Adrian J.  Barker}
	\affiliation{Department of Applied Mathematics, School of Mathematics, University of Leeds, Leeds LS2 9JT, United Kingdom}
	\author{Frank Stefani}
	\affiliation{Helmholtz-Zentrum Dresden-Rossendorf, Bautzner Landstr. 400, D-01328 Dresden, Germany}

	\date{\today}
	\begin{abstract}
		
		We demonstrate, by studying precession-driven flows, that inertial wave hydrodynamic turbulence  can drive a robust magnetic dynamo action. Motivated by the stronger damping of large-scale geostrophic vortices in rapidly rotating planetary and stellar interiors, we introduce a controlled damping of the vortices, which usually accompany inertial wave turbulence and feed on wave energy. It is shown that even a small vortex damping results in a significant increase of the growth rate of the dynamo due to inertial waves in the kinematic regime, allowing it to persist for magnetic Prandtl numbers as low as $Pm \sim 10^{-3}$ and Poincar\'e numbers $Po\sim 0.025$. These critical values of $Po$ and $Pm$ for the dynamo onset decrease with increasing Reynolds number. The onset and growth of the dynamo appear to correlate with the coherent fluctuations of kinetic helicity. Spectral analysis shows that magnetic energy growth is primarily due to inertial-wave-induced induction over a broad range of scales. These results establish inertial waves as an efficient mechanism for magnetic field amplification in rapidly rotating low-$\Pm$ flows relevant to planetary and stellar interiors.
		
	\end{abstract}
	
	\maketitle
	
	Rotating turbulent flows are ubiquitous in planetary interiors, stellar convection zones, oceans, and atmospheres. Their dynamics is governed by the nonlinear interaction between three-dimensional (3D) inertial waves and quasi-two-dimensional (2D) geostrophic vortices \cite{Smith1999, Buzzicotti2018,Alexakis2018, pizzi2022interplay}. Inertial waves can be excited by mechanical forcings such as precession and tides through, respectively, precessional or elliptical instabilities \cite{Kerswell1993,Kerswell2002}. 
	Generally, for fast enough rotation, nonlinear interactions between inertial waves and vortices vanish and hence they decouple \cite{Smith1999}.
	This regime is the most relevant and important one from the astrophysical and geophysical point of view due to very fast rotation of
	stars and planets. 
	Furthermore, in the non-convective and protoplanetary context, the geostrophic modes are most likely to be affected by boundary effects and domain geometry due to their large-scale nature, which may result in their suppression. Another factor favouring inertial waves is that at very weak precession or tidal forcing (responsible for elliptical instability) and small Ekman numbers (i.e., strong rotations) typical of planetary cores, a regime of turbulence dominated by inertial waves rather than by the geostrophic modes may occur \cite{LeReun2017,LeReun_LeBars2020}.  Furthermore, inertial waves excited by tidal forcing play an important role in tidal dissipation in stars and giant planets \cite{Ogilvie_Lin_2004ApJ,Barker_2022ApJL}. 
	In addition, large-scale magnetic fields can also damp the geostrophic  modes \cite{Barker2016}. 
	
	The  capability of hydrodynamic turbulence driven by either elliptical or precessional instability to drive dynamo action in the interiors of planets has been of great interest as a possible source of planetary magnetic fields \cite{Malkus1968, vanyo1991geodynamo, lagrange2011precessional, Barker2013, moffatt_dormy_2019, Rincon2019}. In this regard, precession-driven flows have been investigated more extensively. Experimental \cite{noir2001experimental,Meunier2008,Goto2007,Herault2019,Kumar2023} and numerical studies \cite{kong2015transition,Marques2015,lin2016,Giesecke2018,Albrecht2018,Cebron2019,Lopez2019,Wu2020,Pizzi2021a,Pizzi2021b} have demonstrated precession-driven dynamos in various parameter regimes \cite{Tilgner2005,Wu2009,Cappanera2016,goepfert2016dynamos,Goepfert2019,Giesecke2018,Cebron2019,Kumar2023,Mason2002,Barker2016,Khlifi2018,pizzi2022interplay}. In particular, it was shown that precessional dynamos can be strongly influenced by large-scale geostrophic vortices \cite{Kumar2024}. However, because vortices and inertial waves interact nonlinearly and are both produced in simulations, their individual contributions to dynamo action remain difficult to isolate. Early mean-field studies suggested that helical inertial waves may generate large-scale magnetic fields \cite{Moffatt_1970,Davidson_2014GeoJI,Davidson_Ranjan_2015GeoJI,Ranjan_etal_2018GeoJI,Ranjan_etal2020GeoJI,Tilgner2005}. Specifically, in a seminal paper, Moffatt \cite{Moffatt_1970} analytically showed that the random superposition of inertial waves that lack reflection symmetry, i.e., have the same sign
	of kinetic helicity, can give rise to  a large-scale dynamo, which
	was confirmed and explored in greater detail in subsequent numerical simulations, mostly  within a mean field approach \cite{Davidson_2014GeoJI,Davidson_Ranjan_2015GeoJI,Ranjan_etal_2018GeoJI,Ranjan_etal2020GeoJI}. 
	However, theoretical analysis using a leading-order Stokes drift approach \cite{Herreman_Lesaffre_2011JFM} argued that 
	a single inertial wave does not provide a sufficiently complex drift to sustain dynamo action, whereas nonlinearly interacting inertial waves can drive, at best, only a very weak dynamo \cite{Vidal_Cebron2026}.
	Simulations often consider the combined effects of waves and geostrophic vortices without distinguishing their individual contributions in the dynamics of a magnetic field \cite{Davidson_Ranjan_2015GeoJI,Ranjan_etal_2018GeoJI,Ranjan_etal2020GeoJI,Hide_Stewartson1972RvGSP}. Thus, whether inertial waves can independently drive dynamo at low magnetic Prandtl and Ekman numbers relevant to planetary and stellar interiors still remains an open question.
	
	In this Letter, we investigate the dynamo capability of inertial wave turbulence using direct numerical simulations. This wave turbulence is considered in a local shearing box model and is energetically supplied by hydrodynamic precessional instability \cite{Barker2016,pizzi2022interplay,Kumar2024}. We show that inertial waves are capable of sustaining an efficient dynamo action over a broad range of scales at significantly lower $Pm$ and $Po$ than previously reported, establishing inertial waves as a key mechanism for magnetic field generation in rotating electrically conducting flows.

	We consider a flow in a local Cartesian coordinate frame $(x,y,z)$ rotating around the vertical $z$-axis and precessing around another tilted $x$-axis with angular velocities $\Omega$ and $\Po\cdot\Omega$,  respectively,  where $\Po$ is the Poincar\'{e} number characterizing the strength of precession. In this frame,  also referred to as the  "mantle frame'' of a precessing planet, Coriolis and Poincar\'{e} forces give rise to a laminar background flow proportional to $\Po$, with a linear shear along the $z$-axis, which varies periodically with time at the rate $\Omega$ \cite{Mason2002,Barker2016,pizzi2022interplay},
	\[
	\boldsymbol{U}_0= -2 \Omega\cdot \Po\cdot  z(\sin(\Omega t),\cos(\Omega t),0).
	\]
	The velocity perturbation $\boldsymbol{u}$ about this precessional base flow and the magnetic field  $\boldsymbol{b}$ are governed by the equations of incompressible MHD, which in the rotating and precessing local frame are \cite{Barker2016,Kumar2024} 
	\begin{multline}\label{eq:ns_pert}
		(\partial_t +\boldsymbol{U}_0 \cdot \boldsymbol{\nabla}+\boldsymbol{u} \cdot \boldsymbol{\nabla})\boldsymbol{u} = - \frac{1}{\rho}\boldsymbol{\nabla}\Pi + \frac{1}{\mu_0\rho}(\boldsymbol{b}\cdot\boldsymbol{\nabla})\boldsymbol{b}+\nu \boldsymbol{\nabla}^{2} \boldsymbol{u}  \\- 2\Omega \boldsymbol{e}_z \times \boldsymbol{u} - 2\Omega \boldsymbol{\varepsilon}(t) \times \boldsymbol{u} +2\Omega u_z \boldsymbol{e}_z \times \boldsymbol{\varepsilon}(t),\end{multline}
	\begin{equation}\label{eq:induction}
		(\partial_t + \boldsymbol{U}_0 \cdot \boldsymbol{\nabla})\boldsymbol{b}= \nabla\times(\boldsymbol{u}\times \boldsymbol{b}) +\eta \boldsymbol{\nabla}^{2} \boldsymbol{b} -2\Omega b_z \boldsymbol{e}_z \times \boldsymbol{\varepsilon}(t), 
	\end{equation}
	\begin{equation}\label{eq:divergence}
		\boldsymbol{\nabla} \cdot \boldsymbol{u}=\boldsymbol{\nabla} \cdot \boldsymbol{b}=0,
	\end{equation}
	where $\rho$ is the constant density, $\Pi$ is the sum of thermal and magnetic pressures, $\nu$ is the kinematic viscosity, $\mu_0$ is the  vacuum permeability and $\eta$ is the magnetic diffusivity. The vector $\boldsymbol{\varepsilon}(t)=\Po(\cos(\Omega t), -\sin(\Omega t),0)$ incorporates the effects of precession on the perturbation dynamics in this local model: Coriolis force due to precession, $-2\Omega \boldsymbol{\varepsilon}(t) \times \boldsymbol{u}$, and the stretching term $2\Omega u_z \boldsymbol{e}_z \times \boldsymbol{\varepsilon}(t)$ due to the shear of the base flow $\boldsymbol{U}_0$ in Eq. (\ref{eq:ns_pert}). These two terms jointly give rise to the hydrodynamic precessional instability that excites inertial waves in the flow \cite{Kerswell1993,Kerswell2002,Sahli2010, Barker2016,pizzi2022interplay}. A similar stretching term for the field $-2\Omega b_z \boldsymbol{e}_z \times \boldsymbol{\varepsilon}(t)$ in Eq. (\ref{eq:induction}) is responsible for the amplification of the horizontal field from the vertical one.  
	
	The flow is considered in a cubic box with length $L_x=L_y=L_z=L$, which represents a small part of a global precessional flow far from its boundaries. In this case, we impose
	the shear-periodic boundary conditions commonly used in the local model of the flow \cite{Barker2016,pizzi2022interplay}. We normalize time by $\Omega^{-1}$, lengths by $L$, velocities by $\Omega L$, magnetic field by $\Omega L (\mu_0\rho)^{1/2}$ and pressure by $\rho L^2\Omega^2$. With this normalization the kinetic and magnetic energy densities are $E_k=\boldsymbol{u}^2/2$ and $E_m=\boldsymbol{b}^2/2$, respectively. Together with $\Po$, the main parameters of the flow are the Reynolds number $\Rey=\Omega L^{2}/\nu$, which is the inverse of the Ekman number $Ek=Re^{-1}$ used for rapidly rotating flows in the planetary context, magnetic Reynolds number $\Rm=\Omega L^{2}/\eta$ and the magnetic Prandtl number   $\Pm=\Rm/\Rey$.  In  planetary and stellar interiors, precession is usually  very weak and hence the $\Po$ can be as small as $\Po\lesssim 10^{-4}$. Similarly small is $\Pm\lesssim 10^{-5}$ in planets and  $\Pm \lesssim 10^{-2} $ in stars, while  $\Rey$ is very high  $\gtrsim10^{15}$ (i.e., $Ek\lesssim 10^{-15}$). Although it is not possible to reach this asymptotic parameter regime  with simulations, in an attempt to approach it, in this study we focus on small 
	$\Po$ and $\Pm$ and on high $\Rey$, as far as permitted by our numerical code. This regime of low $\Po$ and high $\Rey$ relevant to planetary cores is, as noted above, the one where inertial wave turbulence with subdominant geostrophic mode is likely to occur \cite{LeReun2017}.
	
\begin{figure}[h]
		\centering
		\includegraphics[width=0.4\columnwidth]{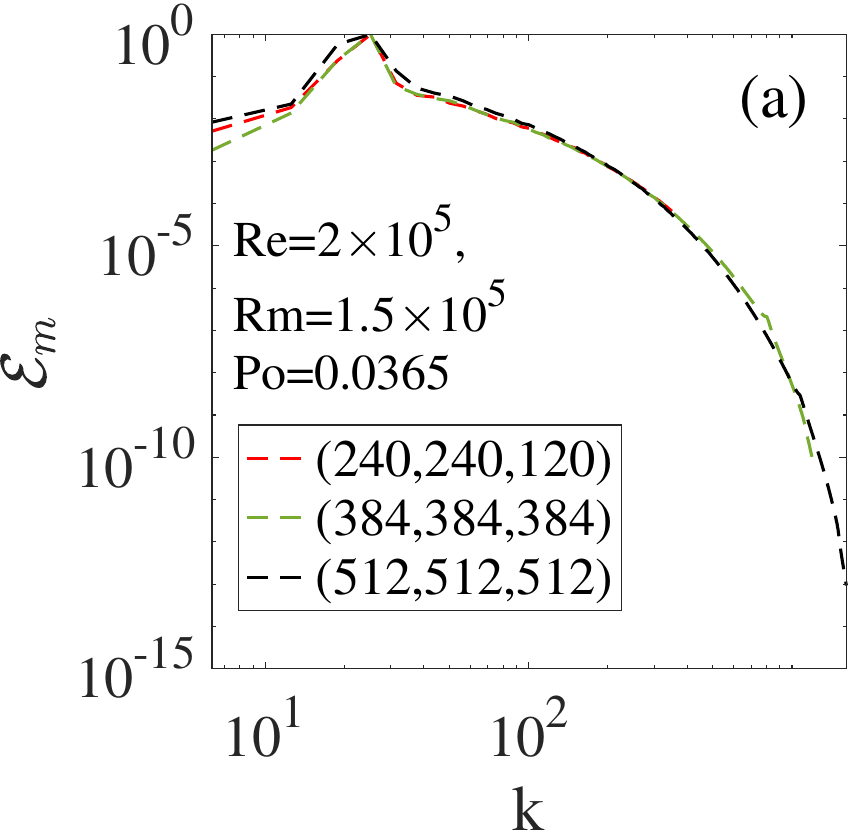}
		\hspace{2em}
		\includegraphics[width=0.4\columnwidth]{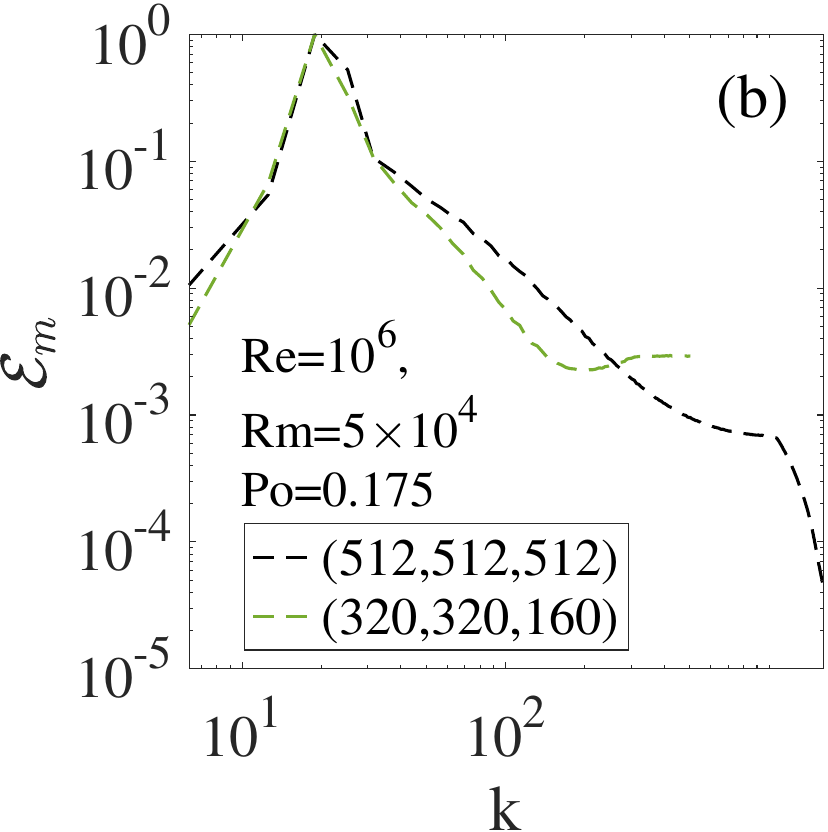}
		\caption{Shell-averaged magnetic energy spectra ${\cal E}_m$ as a function of wavenumber $k$, normalized by their corresponding maxima, are shown for different spatial resolutions in low- and high-$\Po$ regimes, demonstrating convergence in both regimes for (a)  $\Rey=2\times10^5$, $\Rm=1.5\times10^5$, $\Po=0.0365$ and (b) $\Rey=10^6$, $\Rm=5\times10^4$, $\Po=0.175$. }
		\label{fig::resolutiontest}
	\end{figure}

	We solve Eqs. (\ref{eq:ns_pert})-(\ref{eq:divergence}) using the spectral code SNOOPY \cite{Lesur2005} adapted to the considered local model of a precessional flow \cite{Barker2016}. Resolution varies from $(N_x,N_y,N_z)=(128,128,128)$ to $(512,512,512)$ depending on the parameter values, and  is detailed in End Matter (see Fig. \ref{fig::resolutiontest}). Simulations are initialized with  solenoidal random velocity and magnetic field perturbations with the rms amplitude of the latter much smaller than the former by $\langle E_m\rangle/\langle E_k\rangle=10^{-10}$, so that the back-reaction of the magnetic field on the flow remains negligible during the early kinematic growth phase of the field, which we mainly focus on in this work.

\begin{figure} 
		\centering
		\includegraphics[width=0.475\columnwidth]{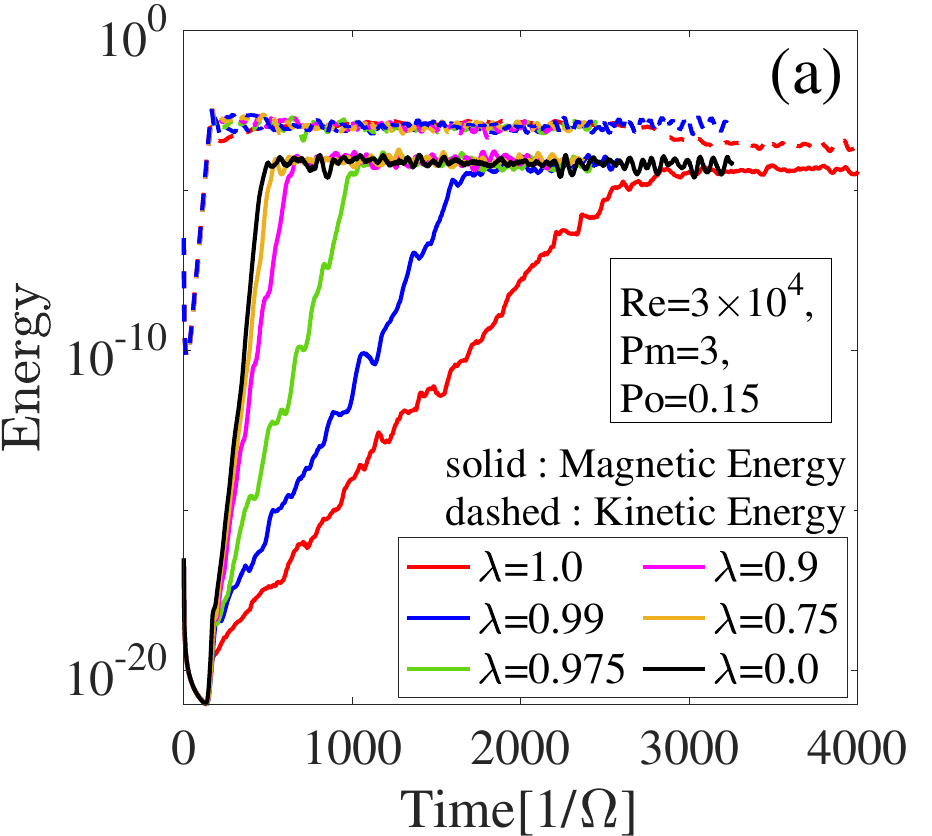}
		\hspace{1em}
		\includegraphics[width=0.45\columnwidth]{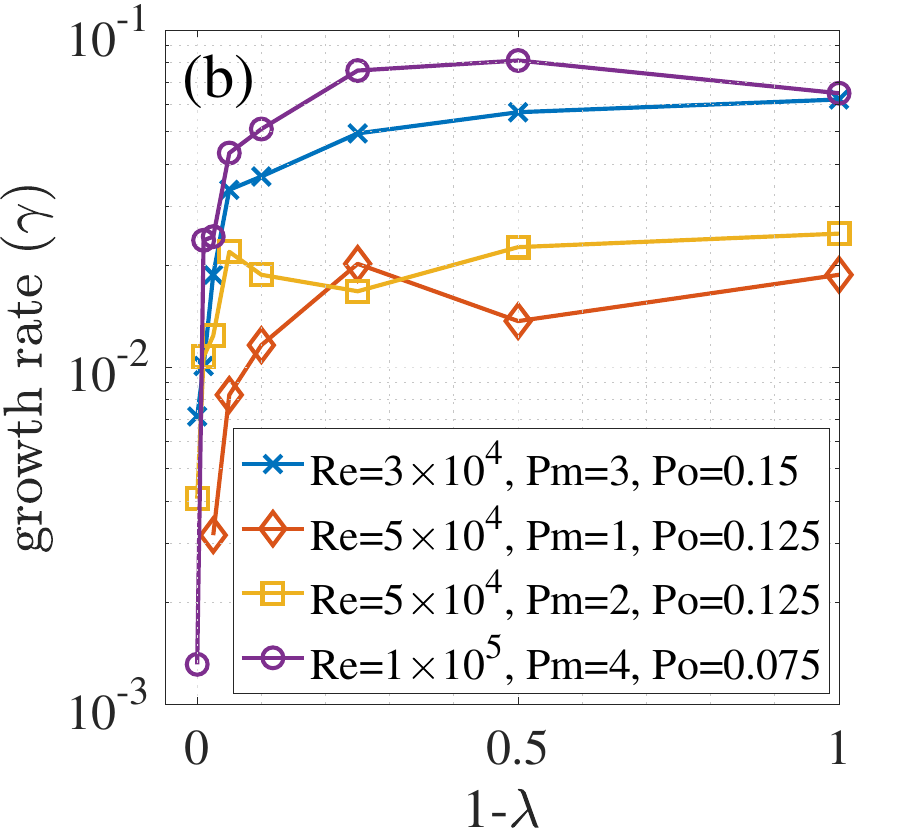}
		\includegraphics[width=0.75\columnwidth, height=0.75\columnwidth]{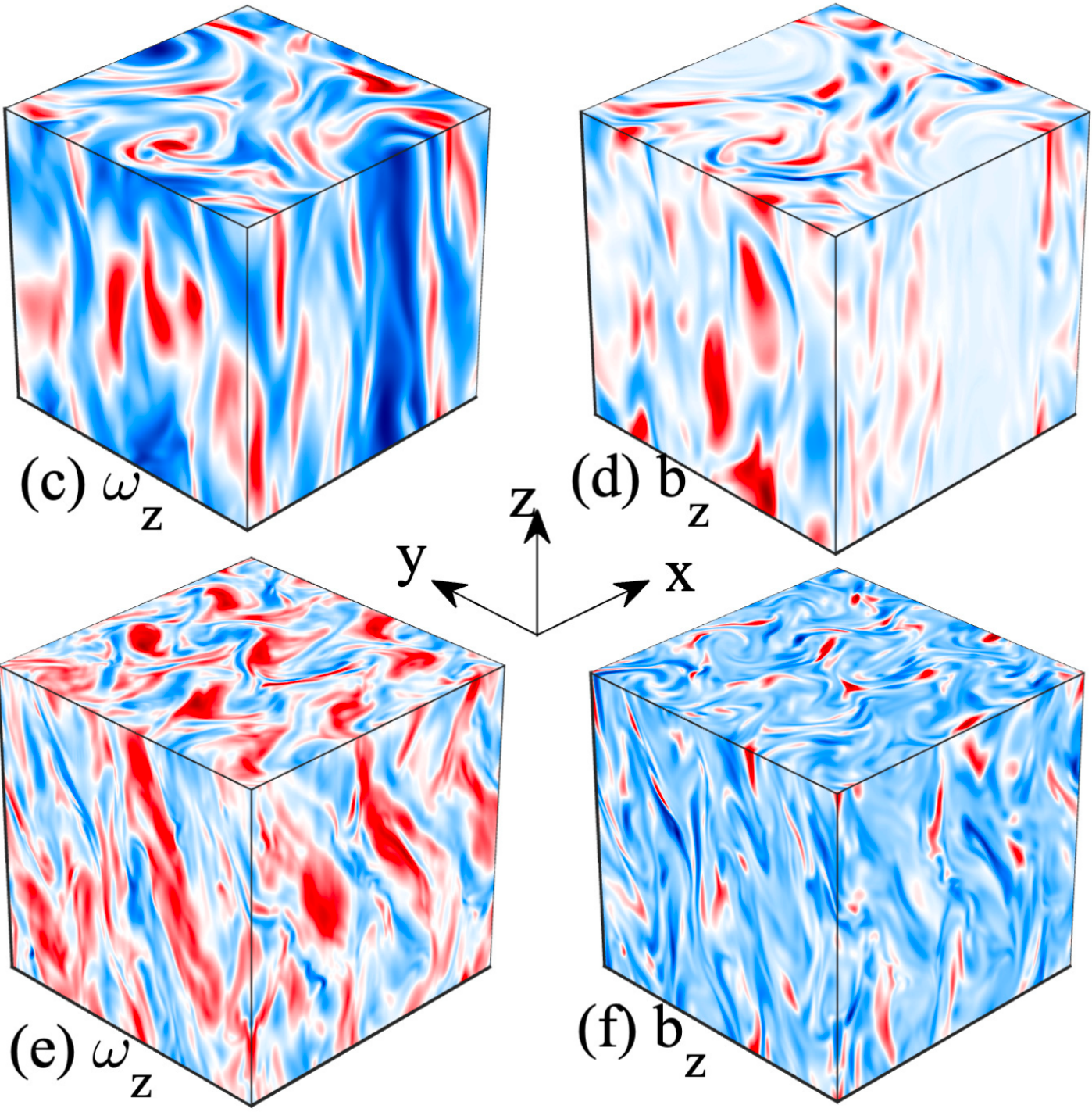}
		\caption{(a) Evolution of the volume-averaged kinetic (dashed) and magnetic (solid) energies for different damping factors $\lambda$ applied to the geostrophic vortices with $k_z=0$ at $\Rey=3\times10^4, Po=0.15$ and $\Pm=3$. (b) Dynamo growth rate $\gamma$ as a function of $1-\lambda$ for different parameters. Distributions of (c,e) the vertical vorticity $\omega_z$ and (d,f) dynamo-induced vertical field $b_z$ in the growth phase in physical space (c,d) with vortices ($\lambda = 1$) and (e,f) without vortices ($\lambda = 0$).}
		\label{fig::energies}
	\end{figure}

	Figure \ref{fig::energies}(a) shows the typical evolution of the volume-averaged kinetic and magnetic energies. 
	In the early linear regime of the precessional instability, the kinetic energy increases exponentially as a result of growing 3D inertial waves, which vary spatially along the vertical $z$-axis with nonzero wavenumber $k_z \neq 0$. After several precession times $(Po\cdot\Omega)^{-1}$, the instability saturates due to nonlinearity into a sustained hydrodynamic turbulence consisting of inertial waves and 2D geostrophic vortices [Fig. \ref{fig::energies}(c)], which are vertically uniform with $k_z=0$. These vortices are sustained by the waves as a result of nonlinear transfers \cite{Barker2016,pizzi2022interplay}. This turbulent state gives rise to dynamo action \cite{Kumar2024},  as is evident from the exponential growth of the magnetic energy in Fig. \ref{fig::energies}(a), which starts just after the saturation of the  precessional instability.  
	
	In rapidly rotating planetary and stellar interior, these large-scale 2D geostrophic vortices are suppressed by boundary effects (e.g., Ekman layers), large-scale magnetic fields or geometric constraints. To mimic the damping of these vortices, following \cite{Le_Reun_2020, LeReun_LeBars2020, Davidson_2014GeoJI}, we introduce a controlled damping parameter $\lambda\in [0,1]$ to isolate the role of inertial waves and quantify their intrinsic dynamo capability.  This damping is applied only to the velocity, $\boldsymbol{u}_{2D}$, of the geostrophic vortical mode at $k_z=0$, such that it is reduced after each time step $\Delta t$ by a factor of $\lambda$, i.e.,  $\boldsymbol{u}_{2D}(t+\Delta  t)=\lambda\boldsymbol{u}_{2D}(t)$. This discrete prescription is analogous to the continuous linear friction term $-f_r Ek^{1/2}\boldsymbol{u}_{2D}$ used in \cite{LeReun2017} for inertial wave turbulence driven by elliptical instability with an effective damping rate $\mu_\lambda=-(1-\lambda)/\Delta t$. Thus,  $\lambda$ is a controlled numerical damping parameter that can model different suppression mechanisms of geostrophic vortices, such as magnetic feedback, geometric constraints or unresolved boundary effects. In this case, $\lambda=0$ corresponds to the complete suppression of the vortices, while for $\lambda=1$ they are fully retained. This allows us to systematically damp the geostrophic modes and isolate the intrinsic dynamo capability of inertial waves in turbulence.
	
	Figure \ref{fig::energies}(a) also shows the evolution of the energies for different $\lambda$. It is seen that the kinetic energy evolution remains largely unchanged across different values of $\lambda$, indicating that the overall development and saturation of the precessional instability are not significantly affected by the damping of vortices. By contrast, the evolution of the magnetic energy appears to be highly sensitive to the presence of vortices: even a slight suppression of vortices with $\lambda=0.99~(\sim 1\%)$ results in the significant increase of the growth rate of the magnetic energy. Suppressing vortices further with $\lambda=0.9 ~(\sim 10\%)$, the magnetic energy grows much faster, approaching growth rates similar to that in the vortex-free ($\lambda=0$) case,  which therefore indicates  a robust dynamo action driven solely by 3D inertial waves. In other words, inertial waves themselves, when not impeded by vortices, can in fact more efficiently drive  a dynamo, even in those regimes where there is no dynamo when vortices are allowed to form in the flow, as seen in Fig. \ref{fig::energies}(b). This figure shows the dynamo growth rate $\gamma=d~{\rm ln}E_m/dt$ as a function of the damping strength $1-\lambda$ for different parameter sets. 
	In all cases, $\gamma$ increases rapidly even for weak damping of the geostrophic vortices, indicating that they strongly inhibit the dynamo, before reaching a plateau for stronger damping. Hence, we focus hereafter on the limiting case of complete vortex suppression ($\lambda=0$), which provides the cleanest framework for isolating and characterizing the intrinsic dynamo capability of inertial-wave turbulence. We then analyze the onset criteria, spectra, and driving mechanism of this dynamo, while the opposite, vortex-dominated regime was investigated in our previous work \cite{Kumar2024}.
	
	Figures \ref{fig::energies}(c)-\ref{fig::energies}(f) show the structures of vertical components of vorticity $\omega_z= (\nabla \times \boldsymbol{u})_z$ and vertical magnetic field $b_z$ in physical space for both cases with and without vortices in the kinematic growth stage. In the first case, the large-scale columnar vortices emerging in the flow [Fig. \ref{fig::energies}(c)] induce the vertical field with similar structure [Fig. \ref{fig::energies}(d)]. This large-scale vortex-driven dynamo in precessional turbulence has been revealed and analyzed  in \cite{Kumar2024}. By contrast,  when vortices are suppressed, the distribution of $\omega_z$ is instead  dominated by small-scale wavy structures due to 3D inertial waves inclined with approximately the same angle to the rotational $z$-axis [Fig. \ref{fig::energies}(e)]. As a result, the magnetic field induced by these waves are also of smaller scale [Fig. \ref{fig::energies}(f)] than those in the presence of vortices. 
	
	\begin{figure} 
		\centering
		\includegraphics[width=0.45\columnwidth]{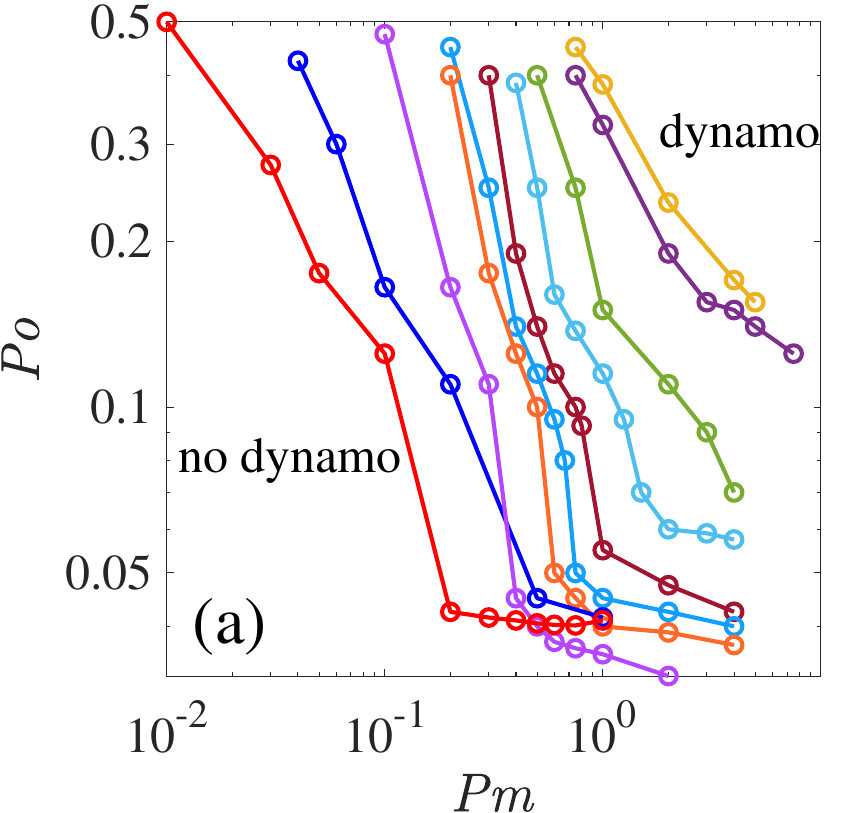}
		\hspace{1em}
		\includegraphics[width=0.45\columnwidth]{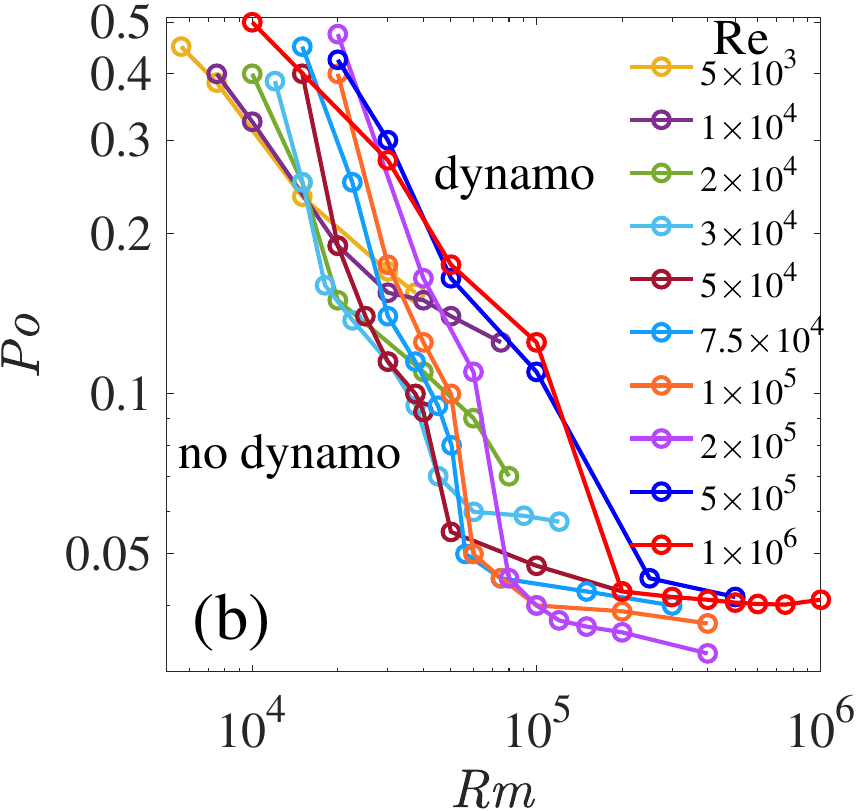}
		\includegraphics[width=0.75\columnwidth, height=0.75\columnwidth]{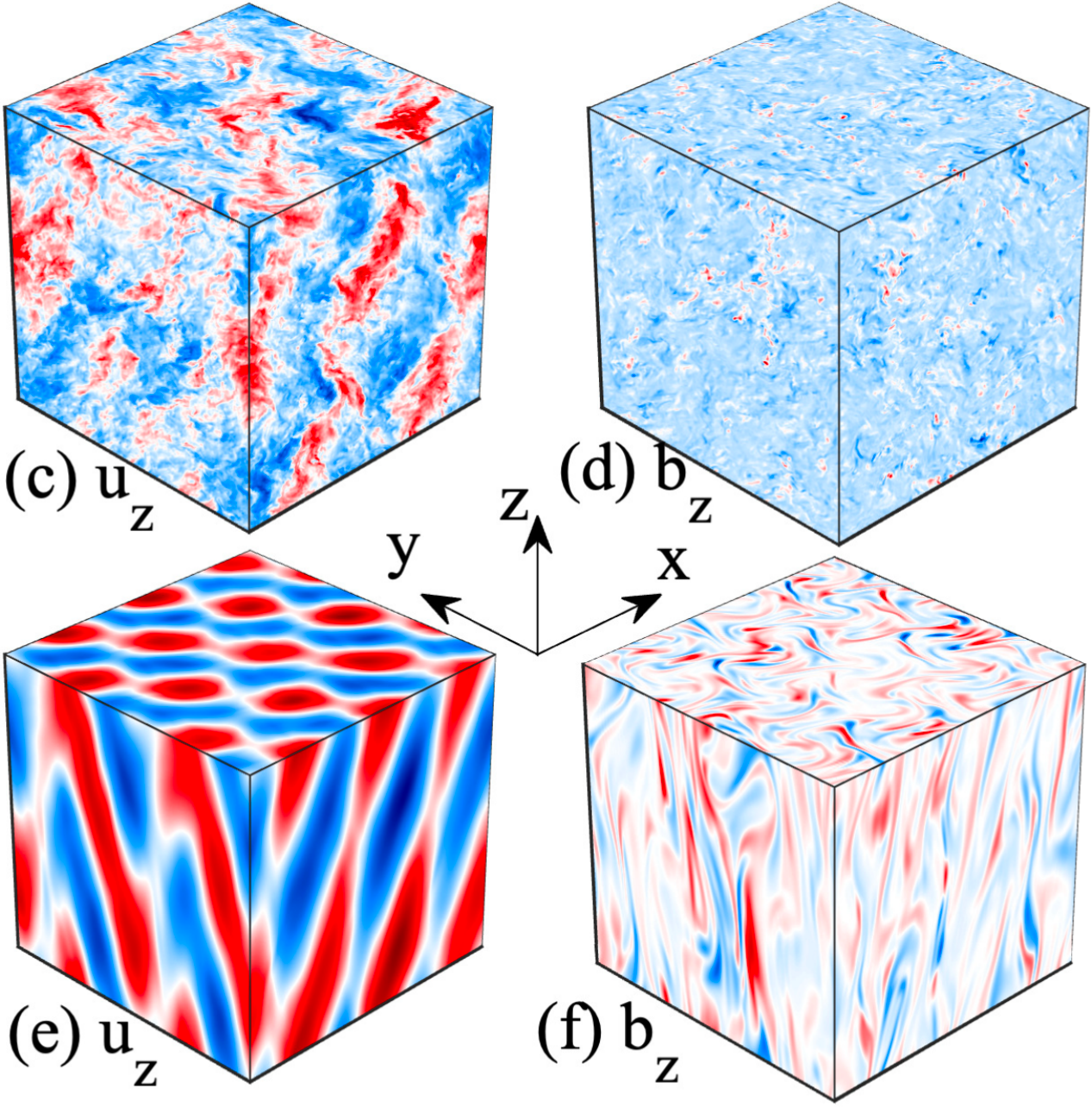}
		\caption{Marginal curves for the dynamo onset at different $\Rey$ in (a) the $(\Pm, \Po)$-plane and (b) in the $(\Rm, \Po)$-plane. The structure of the vertical velocity $u_z$ and magnetic field $b_z$ for (c,d) $\Rm=2\times10^4$ ($\Pm=0.2$) and $\Po=0.45$ and (e,f) for $ \Rm=4\times10^5$ ($\Pm=4$) and $\Po=0.04$ both at $\Rey=10^5$ and $\lambda=0$.}
		\label{fig::scalings}
	\end{figure}
	
\begin{figure} 
		\centering
		\includegraphics[width=\columnwidth]{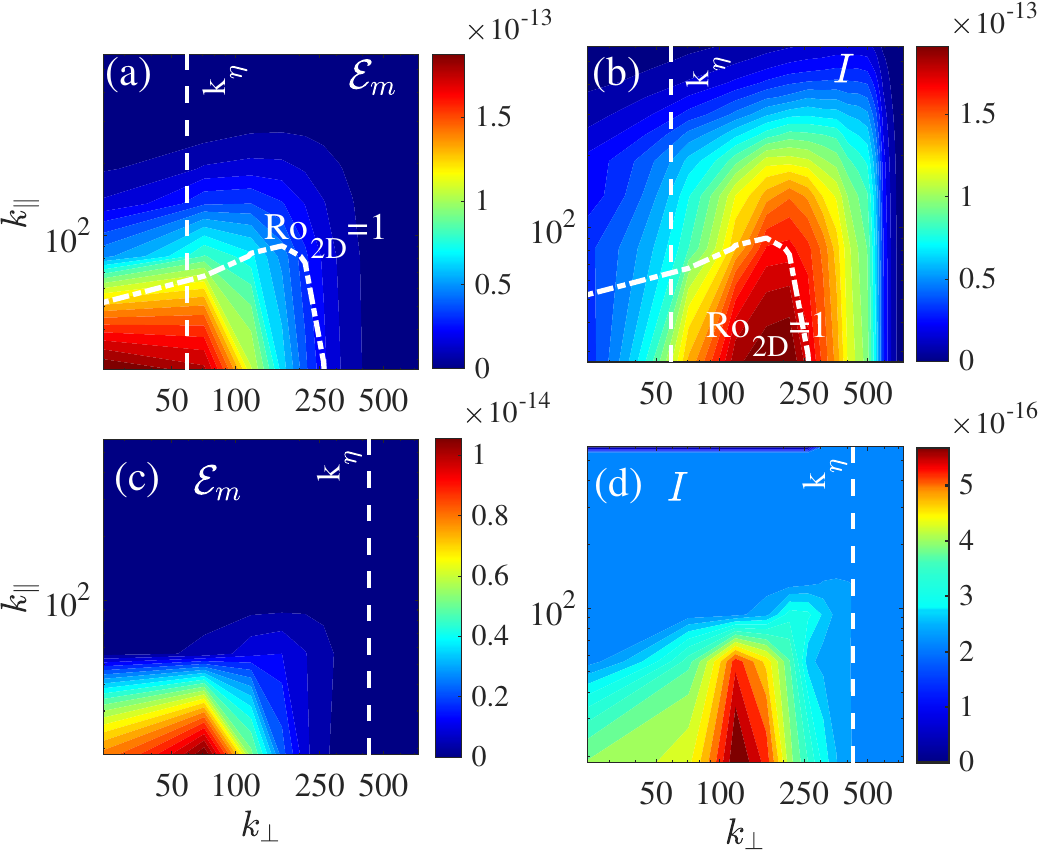}
		\caption{Spectra  of (a,c) the magnetic energy ${\cal E}_m$ and (b,d) the induction term $I$ in Eq. \ref{eq:magnetic_energy_fourier} in the  $(k_\perp,k_\parallel)$-plane during the growth phase of the dynamo with $\lambda=0$ and $\Rey=10^5$ for (a,b) $\Rm=2\times10^4$ ($\Pm=0.2$), $\Po=0.45$ and (c,d) $\Rm=4\times10^5$ ($\Pm=4$), $\Po=0.04$. 
		2D Rossby number $Ro_{2D}=1$ (see text) is also plotted in (a,b) to show the transition between the rotation-dominated inertial wave turbulence at $Ro_{2D} < 1$ and more isotropic 3D turbulence at $Ro_{2D}>1$, where rotation is subdominant. Note that at lower $\Po$ in (c,d), $Ro_{2D}<1$ in the whole range of $k_{\perp}$ and $k_{\parallel}$ due to the dominance of rotation and hence the $Ro_{2D}=1$ line is absent.}
		\label{fig::spectra}
	\end{figure}

	For $\lambda=0$, Fig. \ref{fig::scalings}(a) shows the marginal curves corresponding to zero growth rate of the field, $\gamma=0$, in the $(\Pm,\Po)$-plane for different $\Rey$, which indicate the onset regimes of the inertial wave dynamo. These curves were obtained by conducting a series of kinematic dynamo simulations at fixed $\Rey$ and $\Pm$ (or $\Rm$), while varying $\Po$. For each point, the growth rate $\gamma$ was calculated and critical $\Po$ at which it becomes zero was determined. Repeating this procedure for different $\Pm$ at fixed $\Rey$ gives the marginal stability curve $\gamma(\Rey,\Pm,\Po)=0$. It is seen in Fig. \ref{fig::scalings}(a) that the marginal curves move towards lower $\Pm$ and $\Po$ with increasing $\Rey$, implying the dynamo can extend to smaller and smaller $\Pm$ and $\Po$ at higher $\Rey$. Two different regimes of the dynamo can be distinguished from these marginal curves: (i) a steeper increase of the critical $\Po$ with decreasing $\Pm$ at lower $\Pm$ and (ii) a plateau at smaller $\Po$ and larger $\Pm$, where the dependence of the critical $\Po$ on $\Pm$ is weaker. Interestingly, when these marginal curves are plotted against $\Rm$ instead of $\Pm$ in Fig.~\ref{fig::scalings}(b),  they tend to collapse, somewhat more so in the plateau region. This implies that the dynamo action at small $\Po$ and high $\Rm$ is mainly determined by $\Rm$ rather than by  $\Pm$ or $\Rey$, while at lower $\Rm$ the dependence on $\Rey$ is stronger, although confined to a narrow region. This behaviour at high $\Po$ and low $\Rm$, or equivalently at $\Pm\lesssim 1$, can be understood as a regime dominated by irregular wave motions in the inertial range, since the resistive scale is larger than the viscous scale \cite{Rincon2019}. In this regime, inertial wave coherence remains limited, since as we will see below (Fig. \ref{fig::spectra}), the wave period is comparable to nonlinear interaction time, and hence magnetic field amplification is mainly driven by small-scale, rapidly varying motions, resulting in a strong sensitivity to $\Rm$. By contrast, in the opposite plateau regime with small $\Po$ and large $\Rm$, the rotation is stronger, inertial waves that drive the dynamo are more coherent and large-scale; hence the dynamo is less sensitive to $\Rm$.

	
	

	This difference in the scales of inertial waves driving the dynamo in these two regimes discussed above is evident from Figs. \ref{fig::scalings}(c)-\ref{fig::scalings}(f). These plots compare the typical structures of the vertical velocity $u_z$ and magnetic field $b_z$ in the two dynamo regimes from Fig. \ref{fig::scalings}(a) -- lower $\Pm=0.2$ with stronger precessional forcing $\Po=0.45$ [Figs. \ref{fig::scalings}(c) and \ref{fig::scalings}(d)], corresponding to a steeper part of the curve, and high $\Pm=4$ with lower $\Po=0.04$ [Figs. \ref{fig::scalings}(e) and \ref{fig::scalings}(f)], corresponding to the plateau -- both at $\Rey=10^5$. Specifically,  in the first regime with higher $\Po$, turbulence is dominated by smaller-scale wave structures with higher amplitudes, consistent with our previous study \cite{pizzi2022interplay}, whereas in the second case with smaller $\Po$, larger-scale inertial waves dominate. This is attributed to the supercriticality of the precessional instability. For smaller $\Po$, closer to the instability onset, viscous damping means only larger scale modes (subject to the resonance condition) can grow \cite{Barker2016, pizzi2022interplay}. Consequently, the length scales of the magnetic field generated by these waves are different, as is seen in Figs. \ref{fig::scalings}(d) and \ref{fig::scalings}(f): $b_z$ is of small scale and isotropic at $\Po=0.45$, while it is of larger scale and elongated along the $z$-axis due to rotation at $\Po=0.04$. These behaviors in both dynamo regimes described here in physical space, become even clearer in the spectral analysis below.
	
\begin{figure}[t!]
		\centering
		\includegraphics[width=\columnwidth]{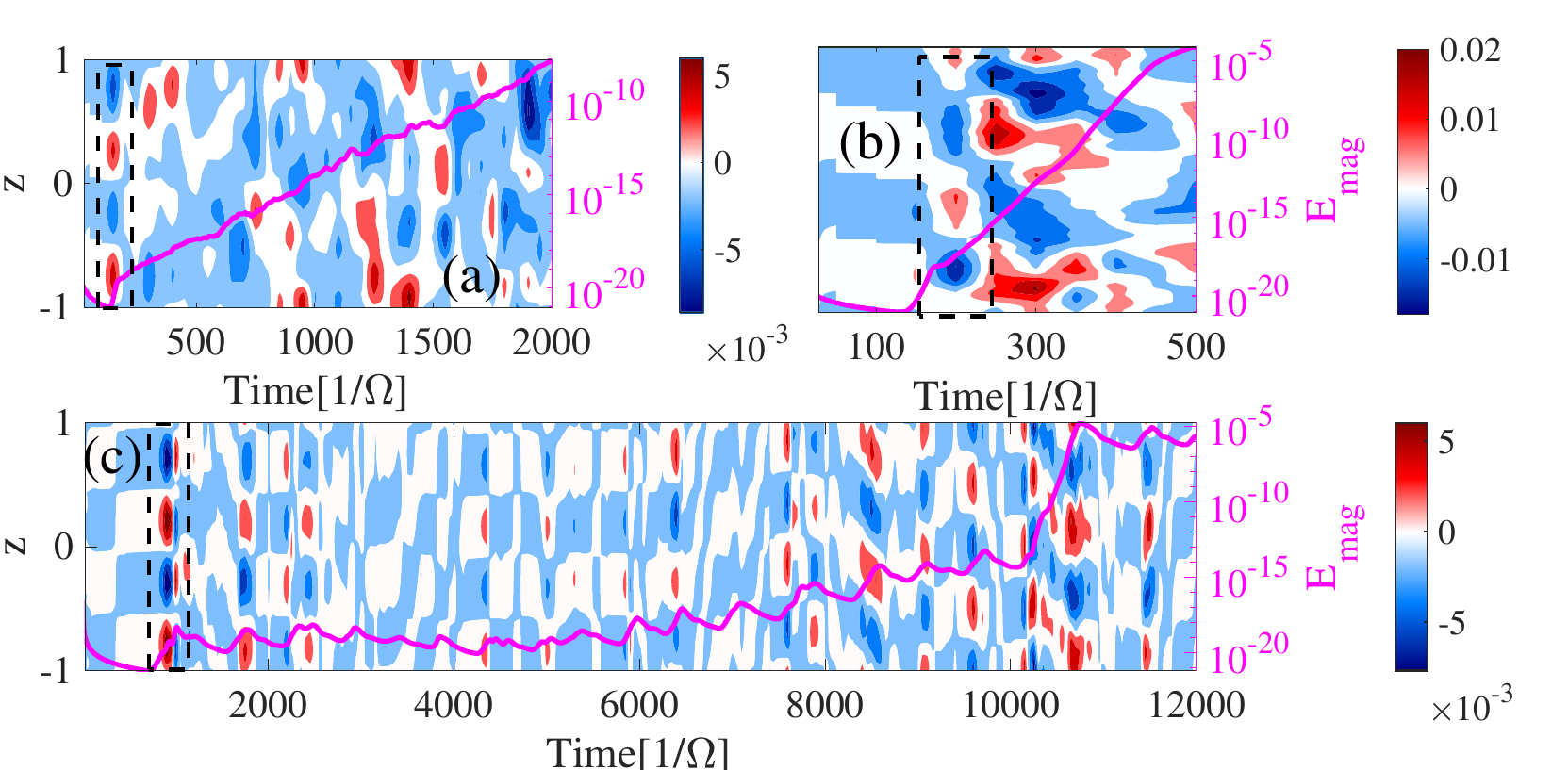}
		\caption{Evolution of the horizontally averaged kinetic helicity $h$ and corresponding magnetic energy (magenta lines) when geostrophic vortices are (a) present ($\lambda=1$) and (b) absent ($\lambda=0$) for the same parameters as in Fig. \ref{fig::energies}(a). (c) Same as (b) but for the parameters in Figs. \ref{fig::scalings}(e) and \ref{fig::scalings}(f). The rectangular dashed box marks the saturation of the precessional instability into turbulence which in turn coincides with the onset of the dynamo.}
		\label{fig::helicity}
	\end{figure}

\begin{figure}[t!]
		\includegraphics[width=0.5\columnwidth]{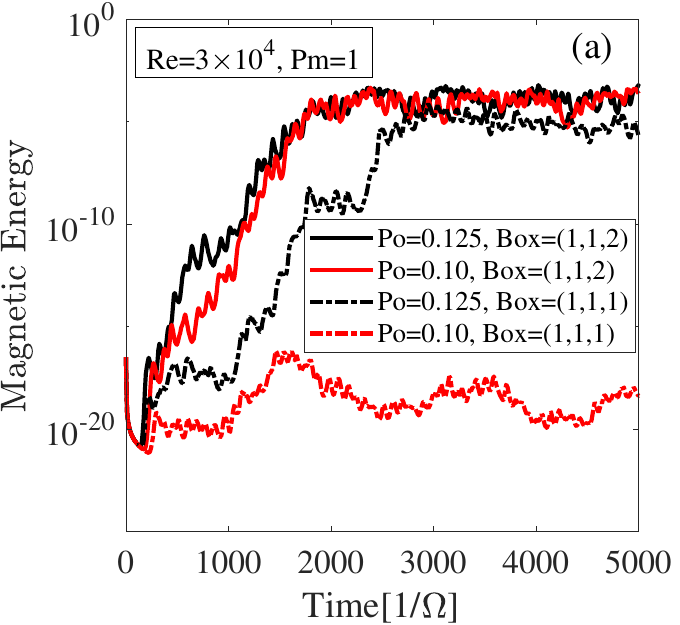}
		\includegraphics[width=0.475\columnwidth]{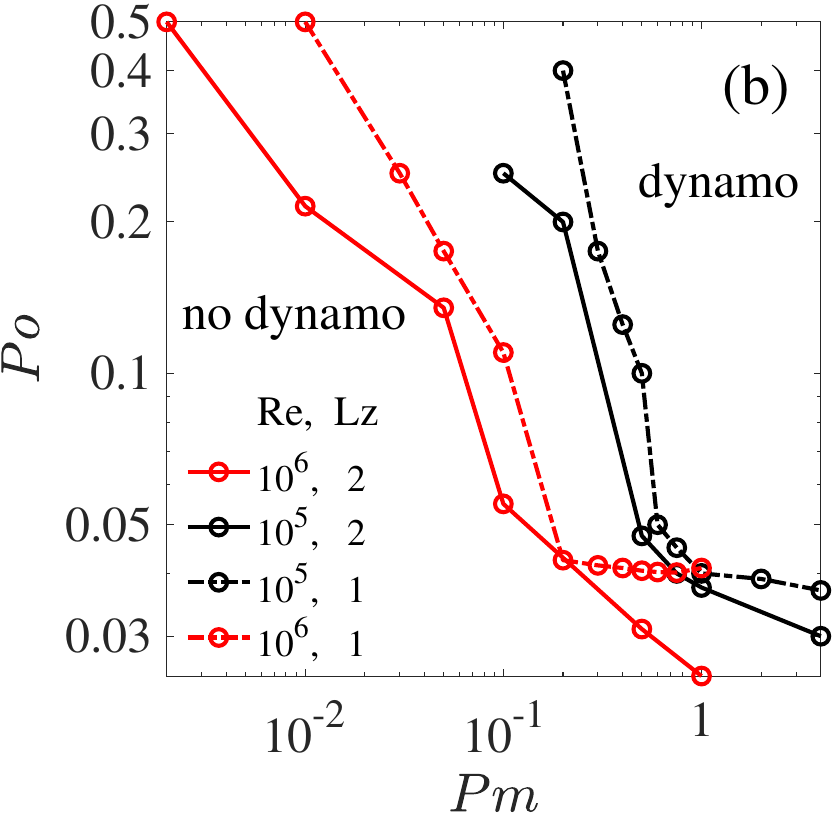}
		\caption{(a) Evolution of the volume-averaged magnetic energies without vortices ($\lambda=0$) for $\Rey=3\times10^4, Pm=1$, $\Po=\{0.1, 0.125\}$ and (b) the marginal curves for the dynamo onset two different box sizes $(L_x, L_y, L_z)=(1,1,1)$ and with the doubled height $(L_x,L_y,L_z)=(1,1,2)$}. 
		
		\label{fig::enegies_boxsize}
	\end{figure}

	Interestingly, the magnetic field growth due to inertial waves in the absence of vortices is strongly correlated with the coherent fluctuations of kinetic helicity emerging after the saturation of the precessional instability (Fig. \ref{fig::helicity} in End Matter), implying that net helical motions associated with inertial waves play a key role in the dynamo dynamics. Moreover, increasing the box size in the vertical $z$-direction enhances the dynamo efficiency with faster growth rate and larger saturation levels (Fig. \ref{fig::enegies_boxsize} in End Matter).

	To better characterize the spatial scales at which dynamo action due to inertial waves is mainly concentrated, we Fourier transform the magnetic field in Eq. (\ref{eq:induction}),  $\boldsymbol{b}(\boldsymbol{r},t)=\int \bar{\boldsymbol{b}}(\boldsymbol{k},t){\rm exp}(i\boldsymbol{k}\cdot \boldsymbol{r})d\boldsymbol{k},$ 
	and the dynamical terms on its right hand side. After multiplying by the complex conjugate $\bar{\boldsymbol{b}}^{\ast}$, we arrive at the equation governing the spectral magnetic energy density ${\cal E}_m=|\bar{\boldsymbol{b}}|^2/2$ \cite{Kumar2024},
	\begin{equation}\label{eq:magnetic_energy_fourier}
		\frac{d{\cal E}_m}{dt} = P + D_M + I,
	\end{equation}
	where $P=\Omega(\bar{\boldsymbol{b}}^{\ast}\bar{b}_z + \bar{\boldsymbol{b}}\bar{b}_z^{\ast})\cdot (\boldsymbol{\varepsilon}(t)\times\boldsymbol{e}_z)$ represents energy exchange of the field with the base flow due to its shear, and $D_m=-2k^2{\cal E}_m/(\Rey\cdot \Pm)<0$ corresponds to resistive dissipation. The term $I= \frac{\mathrm{i}}{2}\left[\bar{\boldsymbol{b}}^{\ast}\cdot\boldsymbol{k}\times(\boldsymbol{u}\times\boldsymbol{b})_{\boldsymbol{k}} - \bar{\boldsymbol{b}}\cdot\boldsymbol{k}\times(\boldsymbol{u}\times\boldsymbol{b})_{\boldsymbol{k}}^{\ast}\right]$ originates from the Fourier transform of the electromotive force $\boldsymbol{u}\times \boldsymbol{b}$ in Eq. (\ref{eq:induction}) and characterizes magnetic energy production due to turbulent velocity. We average these 3D spectra in 2D shells (rings) in the $(k_x,k_y)$-plane and represent them as a function of the perpendicular $k_{\perp}=(k_x^2+k_y^2)^{1/2}$ and parallel $k_{\parallel}=k_z$ (to the rotation axis) wavenumbers.
	
	Figure \ref{fig::spectra} shows the spectra of the magnetic energy ${\cal E}_m$ and the induction term $I$ in the $(k_{\perp},k_{\parallel})$-plane during the growth stage of the dynamo. The term $P$ describing the contribution of the base flow shear in driving the dynamo is by more than an order of magnitude smaller compared to $I$ over all wavenumbers and is therefore not shown here (see Fig. \ref{fig::shellavgspectra} in End Matter). In other words, induction due to turbulent motions rather than the background precessional shear flow  causes the dynamo growth. The induction term $I$ and hence the energy spectra are notably anisotropic in the $(k_{\perp},k_{\parallel})$-plane. For reference, the resistive wavenumber, $k_{\eta}\sim Pm^{3/4}k_{\nu}$ \cite{Rincon2019}, is also shown as a dashed line in the spectra, where  $k_{\nu}\sim (\epsilon/\nu^3)^{1/4}$ is the viscous wavenumber with $\epsilon=\nu\langle(\nabla \times \boldsymbol{ u})^2\rangle$ being the volume-averaged dissipation rate of the kinetic energy.
	For the low-$\Rm$, high-$\Po$ regime [Figs. \ref{fig::spectra}(a) and \ref{fig::spectra}(b)] $I$ spans a wide range in Fourier space and is mostly concentrated at $k_{\perp}\gtrsim k_{\eta}$, that is, the dynamo operates in the inertial range of turbulence, where the velocity field is ``rough'', and hence is sensitive to $\Rm$ in this regime of low $\Rm$, or equivalently low $\Pm \lesssim 1$, as typical of small scale dynamo \cite{Schekochihin2007,Rincon2019}. As a result, ${\cal E}_m$ extends to higher $k_\parallel$ and $k_\perp$, consistent with the predominance of small-scale, irregular wave structures in physical space, as seen in Figs. \ref{fig::scalings}(c) and \ref{fig::scalings}(d). By contrast, for the large-$\Rm$, low-$\Po$ regime [Figs. \ref{fig::spectra}(c) and \ref{fig::spectra}(d)], both ${\cal E}_m$  and $I$ spectra are somewhat more anisotrophic and localized in the $(k_\parallel, k_\perp)$-plane, with a clear concentration toward wavenumbers smaller than the resistive one, $k_{\perp}\lesssim k_{\eta}$. This reflects the dominance of larger-scale, more coherent inertial waves in driving the dynamo consistent with those observed in physical space in Figs. \ref{fig::scalings}(e) and \ref{fig::scalings}(f).
	
	We introduce the 2D Rossby number $Ro_{2D}(k_\perp, k_\parallel)=k_\perp[2\mathcal{E}_k(k_\perp, k_\parallel)]^{1/2}/2\Omega$ to further characterize the  scale-dependent transition in the dynamo-driving process, from a rotation-dominated inertial wave turbulence regime at lower wavenumbers, where $Ro_{2D}<1$, to a relatively isotropic 3D turbulence regime at higher wavenumbers, where $Ro_{2D}>1$, and nonlinear interactions dominate over rotational effects \cite{LeReun2017}. The curve $Ro_{2D}=1$ separating these two regimes is also shown in Fig. \ref{fig::spectra}. As noted above, the induction $I$ spans a broad range of wavenumbers in the $(k_{\perp},k_{\parallel})$-plane. Coherent inertial waves contribute in this term predominantly below the $Ro_{2D}=1$ curve, whereas at higher wavenumbers beyond this curve, at $Ro_{2D}>1$, the contribution comes mainly from faster smaller-scale motions. Note in Figs. \ref{fig::spectra}(c) and \ref{fig::spectra}(d) that in the case of small $Po$,  $Ro_{2D}<1$ at all considered $k_{\perp}$ and $k_{\parallel}$ (hence the $Ro_{2D}=1$ curve is outside the given range of $k_{\perp}$ and $k_{\parallel}$ and not visible in these plots), indicating the dominance of inertial waves over smaller scale irregular motions due to nonlinear interactions in the dynamo dynamics. This is also confirmed by the shell-averaged spectrum of $Ro$ in Figs. \ref{fig::rossbyshellavgspectra}(a) and  \ref{fig::rossbyshellavgspectra}(b),  illustrating that this number is less than  unity at all $k$.

	In this Letter, we have demonstrated, based on the example of precession-driven inertial wave turbulence, that inertial waves are capable of sustaining magnetic dynamo action over a broad range of scales.  When not drained by energy transfers to 2D geostrophic vortical modes, inertial waves can sustain a dynamo at relatively weak precessional forcing amplitudes with thresholds as small as $\Po \sim 0.025 $ and values of $\Pm \sim 10^{-3}$. At small $\Po$ and large $\Rm$ -- a parameter regime relevant to astrophysical bodies -- the dynamo is driven mainly by relatively large-scale, coherent inertial waves. In the non-convective, weakly precessing (with $\Po \lesssim 10^{-4}$) liquid cores of such bodies, inertial wave turbulence may prevail over 2D geostrophic modes and hence play an important role in the magnetic field generation in those bodies. While at large $\Po$ and smaller $\Rm$, the dynamo operates at small scales, it is still driven mainly by inertial waves. At low $\Rm$, but higher $\Po$, small-scale dynamics dominate, leading to a steep dependence of dynamo onset and dynamics on $\Rm$, whereas at large $\Rm$, but small $\Po$, the flow becomes more organized and anisotropic, resulting in an asymptotic behaviour weakly depending on $\Rm$. 
	
	Given that astrophysical flows in stellar and planetary interiors typically operate at orders of magnitude larger Reynolds numbers than those considered here, {i.e., $\Rey \gg 10^5$}, our results broadly suggest that even a weak precessional forcing, or generally, any mechanism exciting a collection (turbulence) of inertial waves, may be sufficient to trigger dynamo in such systems. Specifically, this work characterizes dynamo action in inertial wave turbulence driven by precessional instability,  but our results can also carry over to other mechanically forced systems such as tidally driven flows \cite{Barker_Lithwick_2014MNRAS437, Astoul_Barker_2025MNRAS541}.
	
	This work is supported by the
	Deutsche Forschungsgemeinschaft (DFG) with Grant
	No. MA10950/1-1 and Shota Rustaveli National Science
	Foundation of Georgia (SRNSFG) (Grant No. FR-23-1277). AJB was supported by STFC grants ST/W000873/1 and UKRI1179.

	\bibliography{biblio}
	\clearpage
	\newpage
	\onecolumngrid
	\begin{center}
		{\bf End Matter}
	\end{center}
	\vspace{1em}
	\twocolumngrid
	
	\vspace{-3em}
	\subsection{Resolution Study and shell-averaged spectra}
	
	As a resolution test, we present in Fig. \ref{fig::resolutiontest} the shell-averaged magnetic energy spectra ${\cal E}_m$, normalized by their corresponding maxima,  as a function of wavenumber magnitude $k=(k_{\perp}^2+k_{\parallel}^2)^{1/2}$ for representative cases in both low- and high-$\Po$ regimes and with different spatial resolutions. This confirms that the dynamo growth are well resolved in our problem. 
	
\begin{figure} 
		\centering
		\includegraphics[width=\columnwidth]{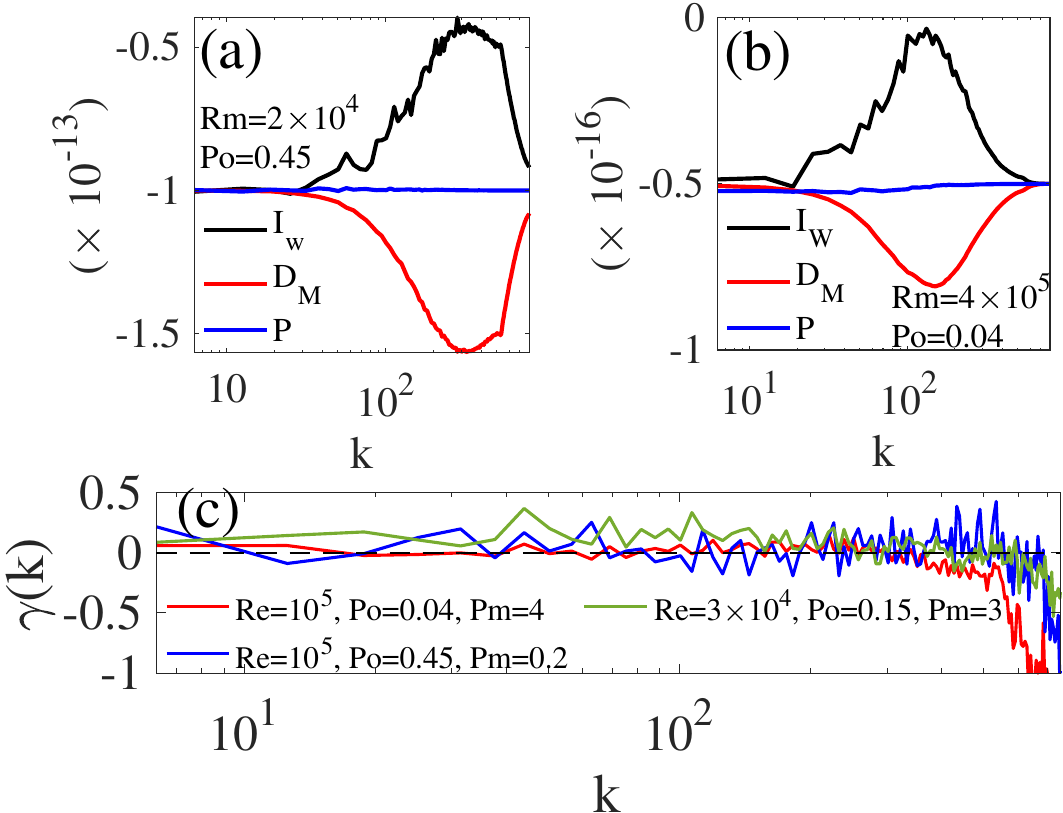}
		\caption{(a) and (b) Shell-averaged spectra of the dynamical terms: induction $I$, magnetic dissipation $D_M$ and shear $P$, in the magnetic energy budget (Eq. 4) for the same parameters as in Fig. \ref{fig::spectra}. (c) The corresponding growth rate $\gamma(k)$ of the spectral magnetic energy vs. $k$ in the marginal case (red and blue) and in the growing case for $\Rey = 3\times 10^4, \Po=0.15 $ and $\Pm =3$ (green) corresponding to that in Fig. 1(a) at $\lambda=0$.}  
		\label{fig::shellavgspectra}
	\end{figure}

	Figures \ref{fig::shellavgspectra}(a) and \ref{fig::shellavgspectra}(b) show the shell-averaged spectra of the induction term $I$ due to waves, the magnetic dissipation $D_M$ and the flow shear term $P$ for the the same parameters as in Fig. \ref{fig::spectra}. In both cases, $I$ dominates in the dynamo driving process over all dynamically relevant scales, while $P$ remains subdominant by more than an order of magnitude. This demonstrates that magnetic field amplification is primarily caused by induction due to motions associated with (inertial wave) turbulence rather than by background shear of the precessional flow.  The resulting growth rate of the dynamo as a function of wavenumbers, $\gamma (k)>0$, are flat with respect to $k$ and extends over a broad range of wavenumbers, indicating that the dynamo operates on all scales. The cut off wavenumber, for which the dynamo is switched off, i.e. $\gamma(k)\leq 0$, increases with $Po$.

\begin{figure} 
		\centering
		\includegraphics[width=\columnwidth]{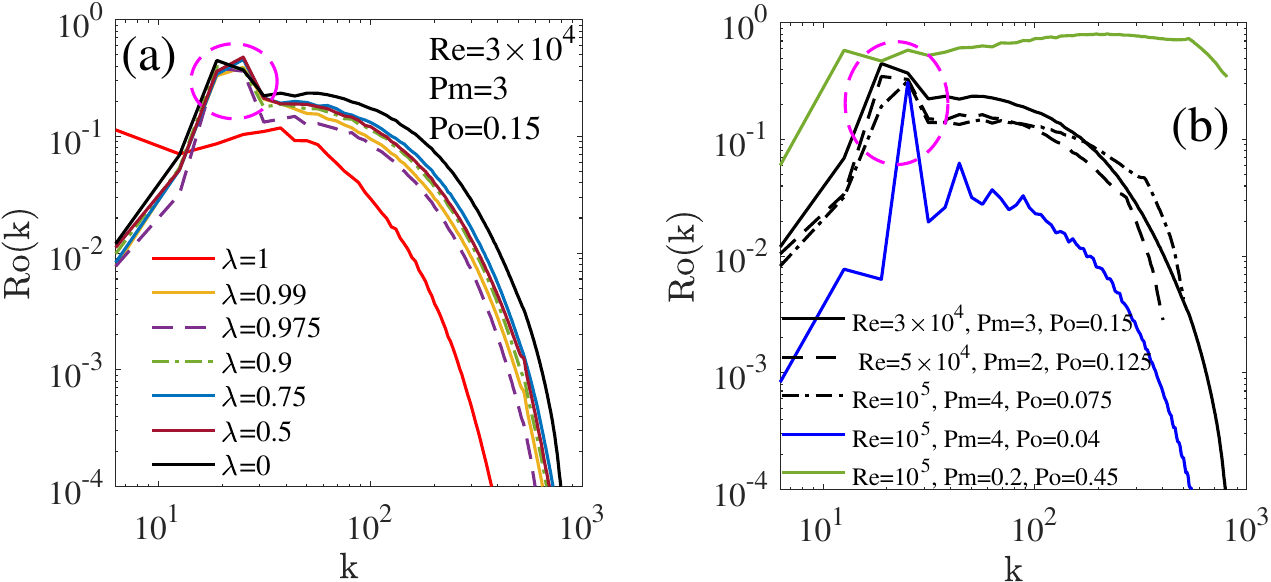}
		\hspace{2em}
		\vspace{-1em}
		\caption{Shell-averaged spectra of the Rossby number $Ro(k)$ vs. $k$ for (a) the same parameters as in Fig. \ref{fig::energies}(a) but different $\lambda$ and (b) fixed $\lambda=0$ but different $\Rey, \Pm$ and $\Po$. Dashed magenta circles in (a,b) show the resonant modes excited due to the precessional instability.} 
		\label{fig::rossbyshellavgspectra}
	\end{figure}

	 Interestingly, the scale-dependent shell-averaged Rossby number, $Ro=k \sqrt{2{\cal E}_\mathrm k(k)}/2\Omega$, shows that even a weak damping of the 2D geostrophic modes, of order 1\% [Fig. \ref{fig::rossbyshellavgspectra}(a)], promotes the coherence and higher amplitude of inertial wave  modes highlighted by the dashed magenta circle. This behaviour is similar to the resonant inertial wave modes reported in \cite{LeReun2017}. In this regime, the flow remains dominated by the effects of rotation, since $Ro(k)<1$, rather than by small-scale, fast motions due to nonlinearity. The same tendency is observed in the case of fully damped geostrophic vortices ($\lambda=0$) for different values of $\Rey, \Pm$ and $\Po$ [Fig. \ref{fig::rossbyshellavgspectra}(b)]. At higher $Po$, the values of $Ro(k)$ extend to higher $k$ and become comparable to 1, indicating the increasing role of nonlinear effects, giving rise to smaller-scale fast 3D motions (see also Fig. \ref{fig::spectra}). 
	
	
	\subsection{Kinetic helicity evolution}
	
	Figures \ref{fig::helicity}(a) and \ref{fig::helicity}(b) compare the evolution of the kinetic helicity density $h = \boldsymbol{u} \cdot (\nabla \times \boldsymbol{u})$ integrated in the $(x,y)$-plane and plotted as a function of time $t$ and vertical coordinate $z$ together with the evolution of the corresponding volume-integrated magnetic energy without ($\lambda=0$) and with ($\lambda=1$) geostrophic vortices, respectively, for the same parameters as in Fig. \ref{fig::energies}(a). 
		Note that in both cases the formation of a certain organized spatio-temporal structure of kinetic helicity just after the saturation of precessional instability (marked by a dashed rectangle) is observed, which also corresponds to the onset of the dynamo, indicating that the dynamo initially sets in due to inertial waves. 
		In the regime where vortices are not suppressed and where they gradually emerge as a result of nonlinear energy transfers from the waves [Fig. \ref{fig::helicity}(a)], traces of organized helicity patterns can still be identified, although they are much less pronounced.
		Thus, vortices tend to make the helicity distribution spatio-temporally more irregular/incoherent and much weaker in magnitude. This in turn results in the reduction of the dynamo growth rate from its highest value just after the dynamo onset, when helicity still has a regular structure before the emergence of the vortices.
		On the other hand, when vortices are suppressed [Fig. \ref{fig::helicity}(b)], the flow is always dominated by inertial waves and exhibits more organized spatio-temporal structure of its helicity, which reflects the intrinsic helical nature of inertial wave turbulence that drives a much faster amplification of the magnetic field ($\approx 7.3$ times) when compared to the first case with vortices in Fig. \ref{fig::helicity}(a). 
		This interpretation is further supported by the volume-integrated kinetic helicity $H=\int hdV$, which is larger when vortices are suppressed, $H=0.0481$, than that in the case with vortices, $H=0.0136$. This larger value of $H$ indicates that the flow possesses a stronger net helicity which is consistent with a Moffatt-type inertial wave dynamo picture \cite{Moffatt_1970}, in which dynamo action is favoured when the wave field has a non-zero helical bias rather than only locally alternating helicity.
		A similar, but somewhat complex behaviour, is observed in Fig. \ref{fig::helicity}(c) for the case without vortices at lower $\Po = 0.04$ and large $\Pm = 4$ [Figs. \ref{fig::scalings}(e) and \ref{fig::scalings}(f)]. In this regime, although kinetic helicity has an overall regular spatio-temporal pattern in the $(t,z)$-plane, there are also intervals exhibiting less regularity. The magnetic field tends to grow mostly during those intervals when helicity is larger with better spatio-temporal organization.
	

	\subsection{Effect of vertical box size $L_z$}
	
	Figure \ref{fig::enegies_boxsize}(a) shows the evolution of the box-averaged magnetic energy with vortices suppressed for two different box sizes: the original one $(L_x, L_y, L_z)= (1,1,1)$ adopted in the text and one with doubled height $(L_x, L_y, L_z)= (1,1,2)$. It is seen that doubling the box size in the vertical $z$-direction causes the dynamo to be more efficient with faster growth rate and larger saturation levels.  
	The increase in $L_z$ provides better scale separation and captures more dynamo-unstable modes. As a result, larger-scale magnetic fields can be generated, which decay more slowly due to reduced Ohmic dissipation and grow faster. The critical $\Pm$ and $\Po$ for the dynamo onset also decrease with increasing $L_z$, as demonstrated by the systematic downward shift of the marginal curves in the $(\Po, \Pm)$-plane as $L_z$ increases from 1 to 2 in Fig. \ref{fig::enegies_boxsize}(b). 
	

\end{document}